\documentclass[3p,12pt]{elsarticle}

\usepackage{amssymb}
\usepackage{amsmath}
\usepackage{amsfonts}
\usepackage{multirow}
\usepackage[group-separator={,}]{siunitx}

\journal{Corrosion Science}

\begin{document}

\begin{frontmatter}

\title{Computational modeling of degradation process of biodegradable magnesium biomaterials}

\author[1]{Mojtaba Barzegari\corref{cor1}}
\ead{mojtaba.barzegari@kuleuven.be}

\author[2,3]{Di Mei}
\ead{di.mei@hzg.de}

\author[2]{Sviatlana V. Lamaka}
\ead{sviatlana.lamaka@hzg.de}

\author[1,4]{Liesbet Geris}
\ead{liesbet.geris@kuleuven.be}
\ead{liesbet.geris@uliege.be}
\ead[url]{www.biomech.ulg.ac.be}

\cortext[cor1]{Corresponding author, Tel: (+32) 16 193831}

\address[1]{Biomechanics Section, Department of Mechanical Engineering, KU Leuven, Leuven, Belgium}
\address[2]{Institute of Surface Science, Helmholtz-Zentrum Geesthacht, Geesthacht, Germany}
\address[3]{School of Materials Science and Engineering \& Henan Key Laboratory of Advanced Magnesium Alloy, Zhengzhou University, Zhengzhou, PR China}
\address[4]{Biomechanics Research Unit, GIGA in Silico Medicine, University of Liege, Liege, Belgium}

\begin{abstract}
Despite the advantages of using biodegradable metals in implant design, their uncontrolled degradation and release remain a challenge in practical applications. A validated computational model of the degradation process can facilitate the tuning of implant biodegradation by changing design properties. In this study, a physicochemical model was developed by deriving a mathematical description of the chemistry of magnesium biodegradation and implementing it in a 3D computational model. The model parameters were calibrated using the experimental data of hydrogen evolution by performing a Bayesian optimization routine. The model was validated by comparing the predicted change of pH in saline and buffered solutions with the experimentally obtained values from corrosion tests, showing maximum 5\% of difference, demonstrating the model’s validity to be used for practical cases.
\end{abstract}

\begin{keyword}
Degradable metals \sep Magnesium corrosion \sep Reaction-diffusion models \sep Finite element method \sep In silico medicine
\end{keyword}

\end{frontmatter}



\section{Introduction}

\subsection{Magnesium biodegradation}

Due to their bio-friendly properties, biodegradable metallic biomaterials, including magnesium (Mg), iron (Fe), and zinc (Zn), are regaining attention in recent years \cite{Zheng2014}. These biomaterials find important applications in the design and manufacturing of supportive implants such as temporary devices in orthopedics and the cardiovascular field  \cite{Chen2014,Zhao2017}. In orthopedics, the biodegradable metallic biomaterials are used as fixation devices, providing adequate support in the early stages while being absorbed gradually during the bone healing process \cite{Qin2019}. Implants fabricated using Mg and its alloys are being used for such a purpose \cite{Riaz2018} due to the similarity of the stiffness between natural bone and Mg, which helps to reduce the stress shielding induced by the implanted device. Additionally, Mg is reported to have a non-toxic contribution to the human body's metabolism and the bone healing process, which makes the release and absorption of metallic ions safe and biocompatible \cite{Xin2008}.

Accumulation of mechanistic understanding of Mg degradation achieved by experimental approaches over the years gradually provided a mechanistic understanding of the biodegradation process. Combining these insights with \textit{in silico} modeling approaches enables researchers to study the biodegradation properties and behavior of the implant in a virtual environment prior to conducting any \textit{in vitro} or \textit{in vivo} tests. When fully validated, computational modeling can (in part) replace certain stages of costly and time-consuming experiments verifying the expected degradation behavior of the designed implants. Additionally, the developed models can be efficiently combined with existing computational models to examine  other related phenomena such as tissue growth or mechanical integrity. 

\subsection{Computational modeling of Mg degradation}

Previous contributions to the computational modeling of the degradation process include a wide range of different approaches, from the basic phenomenological implementations to comprehensive mechanistic models that take into account various aspects of the degradation and resorption process. 


Continuous damage (CD) modeling has always been a common approach for corrosion simulation, but from a physicochemical point of view, it focuses on the mechanical integrity of the degradation and neglects the diffusion process. As a result, its application in the degradation modeling of biomaterials, which includes various fundamental phenomena such as mass transfer through diffusion and reaction, is relatively limited. Despite this issue, a CD model proposed by Gastaldi et al. showed a good performance for simulation of bioresorbable Mg-based medical devices \cite{Gastaldi2011},
in which geometrical discontinuities were interpreted as the reduction of material.

Alternatively, mathematical modeling using transport phenomena equations has shown great flexibility in capturing different mechanisms involved in the biodegradation process. 
As an example, in Ahmed et al., a set of mathematical equations in cylindrical and spherical coordinates was derived to model the chemical reactions of Mg degradation \cite{Ahmed2017}. 
Despite the simplicity of their approach from the computational perspective, their model was able to demonstrate the contribution of various chemical components to the \textit{in vitro} degradation of Mg. Similarly, Grogan et al. developed a mathematical model based on the Stefan problem formulation in 1D space to correlate the mass flux of metallic ions into the solution to the velocity of shrinkage of the material during degradation \cite{Grogan2014}. This was done by considering the mass diffusion and change of the concentration of $\mathrm{Mg}^{2+}$ ions, and then, employing an arbitrary Eulerian-Lagrangian (ALE) approach to extend the model to 3D on an adaptive mesh. 
A similar approach was taken by Shen et al. to develop a theoretical model of the degradation behavior of Mg-based orthopedic implants showing great consistency with \textit{in vitro} test results \cite{Shen2019}. 

An ultimate application of the computational modeling of the biodegradation process of biomaterials can be the prediction of how biodegradation affects the shape of the bulk material, medical device, or implant over time. One of the ways to achieve such a prediction is to capture  the movement of the corrosion front mathematically using an appropriate method. The level set method (LSM) is a widely used example in this regard, which is an implicit mathematical way of representing the moving interfaces. This approach was used in Wilder et al. to study galvanic corrosion of metals \cite{Wilder2014}. They employed LSM on an adaptive mesh to track the moving corrosion interface, but  their model lacked a thorough validation using experimental data. 
Gartzke et al. also worked on a simplified representation of the interface movement by developing a mechanochemical model of the biodegradation process, which helped them to study the effect of degradation on the mechanical properties \cite{Gartzke2020}. They performed a basic qualitative validation on the predictions made by the model.
Another similar study in this regard is the Sun et al. work \cite{Sun2012}, in which a detailed mathematical model was derived and validated to study the deposition of  corrosion products on the surface of materials. 
This mathematical approach was also employed in the biomedical field by Bajger et al. to study the mass loss of Mg biomaterials during biodegradation \cite{Bajger2016}. They used LSM as well as a set of reaction-diffusion equations to track the change of geometry, which can be directly correlated to the loss of material. The derived equations were also able to capture the formation of the corrosion film that decreases the rate of degradation. 
Another comprehensive mathematical model was developed by Sanz-Herreraa et al. to investigate the role of multiple chemical components involved in the \textit{in vitro} degradation of Mg implants \cite{Sanz-Herrera2018}. 
One important drawback of this study was its 2D nature. Although the computational model was capable of studying the effect of multiple components, due to the high number of derived equations, it would be difficult to extend and use the same model for real 3D implants. Additionally, a 2D model cannot capture the full phenomenon of corrosion, and as a result, the validation of the model will be more qualitative. It was shown in the study conducted by Gao et al. \cite{Gao2018}, where they compared the results of a multi-dimension model of the degradation of cardiovascular stents with those of a single-dimension model, that the number of considered dimensions had an important effect on the model predictions. In the end, it is worth mentioning that no dedicated experiments were performed in the aforementioned studies to validate the constructed mathematical and computational models.

\subsection{Objective of current contribution}

The current study focuses on developing a physicochemical model of the biodegradation process of commercially-pure (CP) Mg biomaterials by continuing the work of Bajger et al. \cite{Bajger2016}. In this model, a set of partial differential equations (PDE) was derived according to the underlying chemistry of biodegradation, described as reaction-diffusion processes taking place at the interface of the biomaterial and its surrounding environment. The formation of a protective layer, effects of the ions in the solution, and the change in the pH due to the corrosion phenomenon were taken into account in the mathematical model. The corresponding computational model was  implemented in a fully parallelized manner. Model calibration and validation were executed using data obtained from the immersion tests performed in saline (NaCl) and simulated body fluid (SBF) solutions.

\section{Background theory}

\subsection{Biodegradation as a reaction-diffusion system}

The biodegradation process can be considered as a reaction-diffusion system \cite{wang2008}, in which the ions are released due to the chemical reactions on the surface, and the released ions diffuse through the surrounding solution and materials. These ions can interact with other ions and form new compounds \cite{Mei2020}. As the reaction-diffusion systems have been studied in science and engineering for a couple of decades, the analogy with a reaction-diffusion system makes it convenient to construct a mathematical model of the biodegradation process based on the well-established transport phenomena equations \cite{Grindrod1996}. From the mathematical perspective, a reaction-diffusion system is expressed by a set of parabolic PDEs that describe the conservation of contributing chemical species in the studied system.

\subsection{Moving boundary - Stefan problems}

Moving boundary problems, also called Stefan problems, are the general class of mathematical problems in which the boundary of the domain should also be calculated in addition to the solution of the other equations \cite{Crank1987}. Coupling the reaction-diffusion system of biodegradation with a moving boundary problem constructs a mathematical model in which the change of the domain geometry due to the material loss can be correlated to the underlying reaction and diffusion processes of corrosion. As the geometry can be determined accurately, this approach provides a way to measure the mass loss directly by computing the change in the volume of the material. In such a system, the moving boundary is the material-solution interface (corrosion front).

For a 1D corrosion diffusion system, the position of the diffusion interface can be determined by \cite{Crank1987}:
\begin{equation} \label{eq:stefan}
s(t)=s_{0}+2 \alpha \sqrt{t},
\end{equation}
in which the $s(t)$ represents the position at any given time, and $s_0$ is the initial interface position. $\alpha$ coefficient can be calculated using:
\begin{equation} \label{eq:stefan_alpha}
\alpha=\frac{[\mathrm{Mg}]_{0}-[\mathrm{Mg}]_{\text {sat }}}{[\mathrm{Mg}]_{\text {sol }}-[\mathrm{Mg}]_{\text {sat }}} \sqrt{\frac{D}{\pi}} \frac{\exp \left(\frac{-\alpha^{2}}{D}\right)}{\operatorname{erfc}\left(\frac{-\alpha}{\sqrt{D}}\right)}
\end{equation}
where $[\mathrm{Mg}]_{\text {sol}}$ is the concentration in the solid bulk (i.e. materials density) and $[\mathrm{Mg}]_{\text {sat}}$ is the concentration at which the material is released to the medium. $[\mathrm{Mg}]_{0}$ represents the initial concentration of the metallic ions in the medium, which is usually zero for most corrosion cases.

Eqs. \ref{eq:stefan} and \ref{eq:stefan_alpha} can be used to simply track the movement of the corrosion front, which is the employed method in studies like the Gorgan et al. work \cite{Grogan2014}, but apparently, the real-world corrosion problems are 3D and much more complex than the described system. 

As will be described later, Eq. \ref{eq:stefan} is used strictly for the first time step of the simulations in low diffusion regimes for calculating the initial velocity of the interface. Generally speaking, a more sophisticated approach, such as the level set method, is required for tracking the interface of complex 3D geometries. 

\subsection{Level set method}

In the current study, the corrosion front is tracked using an implicit function such that the zero iso-contour of the function represents the metal-solution interface. As a common practice, this implicit function is expressed as a signed distance function that defines the distance of each point of space (the domain of interest) to the interface. Such a definition implies that the zero iso-contour of the function belongs to the interface. The level set method provides an equation to declare such an implicit function, $\phi=\phi(\mathbf{x},t), \mathbf{x} \in \Omega \subset \mathbb{R}^{3}$, which can be obtained by solving \cite{RonaldFedkiw2002}:
\begin{equation} \label{eq:lsm_full}
\frac{\partial \phi}{\partial t}+{\overrightarrow{V^\mathrm{E}} \cdot \nabla \phi}+{\mathrm{V}^\mathrm{N}|\nabla \phi|}={b \kappa|\nabla \phi|}
\end{equation}
in which $\overrightarrow{V^\mathrm{E}}$ is the external velocity field, and  $\mathrm{V}^\mathrm{N}$ is the value of the normal interface velocity. The last term is related to the curvature-dependent interface movement and is omitted. As the effect of perfusion is neglected in the current study, the term containing the external velocity is also eliminated, resulting in the following simplified form of the level set equation:
\begin{equation} \label{eq:lsm_simplified}
\frac{\partial \phi}{\partial t}+\mathrm{V}^\mathrm{N}|\nabla \phi|=0.
\end{equation}

By having the normal velocity of the interface ($\mathrm{V}^\mathrm{N}$) at each point and solving Eq. \ref{eq:lsm_simplified}, the interface can be captured at the zero iso-contour of the $\phi$ function.

\section{Materials and methods}

\subsection{Underlying chemistry}

The chemistry of biodegradation of Mg depends considerably on the surrounding solution and the presence of certain ions \cite{Mei2020}. In NaCl solutions, the anodic and cathodic reactions as well as the formation and elimination of side corrosion products can be considered as follows \cite{Zheng2014}:
\begin{equation} \label{eq:oxidation_react}
\mathrm{Mg}+2 \mathrm{H}_{2} \mathrm{O} \stackrel{k_{1}}{\rightarrow} \mathrm{Mg}^{2+}+\mathrm{H}_{2}+2 \mathrm{OH}^{-} \stackrel{k_{1}}{\rightarrow} \mathrm{Mg}(\mathrm{OH})_{2}+\mathrm{H}_{2}
\end{equation}
\begin{equation} \label{eq:break_react}
\mathrm{Mg}(\mathrm{OH})_{2}+2 \mathrm{Cl}^{-} \stackrel{k_{2}}{\rightarrow} \mathrm{Mg}^{2+}+2 \mathrm{Cl}^{-}+2 \mathrm{OH}^{-}.
\end{equation}

Reaction \ref{eq:break_react} is not fully correct from the chemical point of view. In fact, Mg surface is always covered by MgO layer, and $\mathrm{Mg}(\mathrm{OH})_{2}$ forms on top of that either at atmospheric conditions or during the immersion. The integrity of this MgO layer is undermined by $\mathrm{Cl}^{-}$ ions, leading to an increase in degradation rate:
\begin{equation} \label{eq:break_react_mgo}
\mathrm{MgO}+ \mathrm{Cl}^{-} + \mathrm{H}_{2} \mathrm{O} \stackrel{k_{2}}{\rightarrow} \mathrm{Mg}^{2+}+ \mathrm{Cl}^{-}+ 2\mathrm{OH}^{-}.
\end{equation}

\begin{figure}[h]
\center \includegraphics[width=10cm]{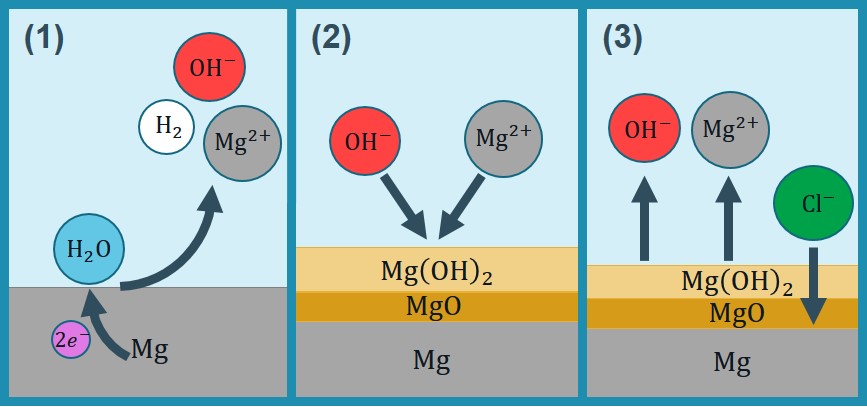}
\caption{The chemistry of biodegradation of Mg considered in the current study: 1) Mg oxidation and water reduction processes accompanied by releasing $\mathrm{Mg}^{2+}$ and $\mathrm{OH}^{-}$  ions as well as $\mathrm{H}_2$ gas, 2) formation of a partially protective precipitation layer, 3) dynamic solubility equilibrium and contribution of $\mathrm{Cl}^{-}$.} \label{fig:chemistry}
\end{figure}

Although $\mathrm{Cl}^{-}$ formally does not participate in reaction \ref{eq:break_react_mgo}, it reflects the dependence of Mg corrosion rate on $\mathrm{Cl}^{-}$ concentration. This effect on the rate of degradation has been widely expressed as the effect of $\mathrm{Cl}^{-}$ on the $\mathrm{Mg}(\mathrm{OH})_{2}$ in the literature \cite{Zheng2014,Zhao2017}. In the developed model, this effect is used interchangeably by omitting the MgO component, so the protective film formed on the corrosion interface is assumed to contain $\mathrm{Mg}(\mathrm{OH})_{2}$ only. Moreover, it has been shown recently that oxygen reduction reaction also takes place during corrosion of Mg \cite{Wang2020,Strebl2020,Silva2018}. However, this is a secondary reaction (complementing water reduction) contributing to 1-20\% of the total cathodic current depending on the conditions. Hence, it is not taken into consideration in this model.  Additionally, the involved chemical reactions are more complicated in SBF solutions due to the presence of further inorganic ions and the formation of a layered precipitate structure  \cite{Mei2020}, but the effect of these ions is currently encapsulated in the reaction rates and the diffusion  coefficients of the developed mathematical model. The summary of the considered chemistry to develop the mathematical model is depicted in Fig. \ref{fig:chemistry}.

\subsection{Mathematical modeling}

To keep track of the concentration changes of various contributing chemical components, we define four state variables for the concentration of $\mathrm{Mg}^{2+}$ ions, protective film ($\mathrm{Mg}(\mathrm{OH})_{2}$), chloride ($\mathrm{Cl}^{-}$) ions, and the hydroxide ($\mathrm{OH}^{-}$) ions:
\begin{equation} \label{eq:state_vars}
\begin{aligned}
&C_{\mathrm{Mg}} = C_{\mathrm{Mg}}(\mathbf{x},t), \quad C_{\mathrm{Film}} = C_{\mathrm{Film}}(\mathbf{x},t)  \\
&C_{\mathrm{Cl}} = C_{\mathrm{Cl}}(\mathbf{x},t), \quad C_{\mathrm{OH}} = C_{\mathrm{OH}}(\mathbf{x},t) \quad \mathbf{x} \in \Omega \subset \mathbb{R}^{3}
\end{aligned},
\end{equation}
which are indeed 4 scalar functions of space and time. $\Omega$ denotes the whole region of interest, including both the Mg bulk and its surrounding medium. By doing this, the value of pH at each point of $\Omega$ can be calculated as:
\begin{equation} \label{eq:ph}
\mathrm{pH} = 14 + \log_{10}{C_{\mathrm{OH}}},
\end{equation}
where $C_{\mathrm{OH}}$ implies the activity of $\mathrm{OH}^{-}$. By having the definition of the state variables in Eq. \ref{eq:state_vars}, the biodegradation of Mg described by Eqs. \ref{eq:oxidation_react} and \ref{eq:break_react} can be represented as a set of reaction-diffusion PDEs:
\begin{equation} \label{eq:pde_mg}
\frac{\partial C_{\mathrm{Mg}}}{\partial t}=\nabla \cdot \left(D_{\mathrm{Mg}}^{e}  \nabla C_{\mathrm{Mg}} \right)-k_{1} C_{\mathrm{Mg}}\left(1-\beta \frac{C_{\mathrm{Film}}}{[\mathrm{Film}]_{\max }}\right) +k_{2} C_{\mathrm{Film}} {C_{\mathrm{Cl}}}^{2}
\end{equation}
\begin{equation} \label{eq:pde_film}
\frac{\partial C_\mathrm{Film}}{\partial t}=k_{1} C_{\mathrm{Mg}}\left(1-\beta \frac{C_{\mathrm{Film}}}{[\mathrm{Film}]_{\max }}\right) -k_{2} C_{\mathrm{Film}} {C_{\mathrm{Cl}}}^{2}
\end{equation}
\begin{equation} \label{eq:pde_cl}
\frac{\partial C_{\mathrm{Cl}}}{\partial t}=\nabla \cdot \left(D_{\mathrm{Cl}}^{e}  \nabla C_{\mathrm{Cl}} \right)
\end{equation}
\begin{equation} \label{eq:pde_oh}
\frac{\partial C_{\mathrm{OH}}}{\partial t}=\nabla \cdot \left(D_{\mathrm{OH}}^{e}  \nabla C_{\mathrm{OH}} \right)+k_{2} C_{\mathrm{Film}} {C_{\mathrm{Cl}}}^{2}
\end{equation}
in which the maximum concentration of the protective film can be calculated according to its porosity ($\epsilon$) \cite{Bajger2016}:
\begin{equation} \label{eq:film_max}
[\mathrm{Film}]_{\max }=\rho_{\mathrm{Mg}(\mathrm{OH})_{2}} \times(1-\epsilon).
\end{equation}
$D^e$ is the effective diffusion coefficient for each component. Due to the formation of the protective film, the diffusion coefficient is not constant and varies from the actual diffusion coefficient of the ions to a certain fraction of it. This fraction can be defined as ${\epsilon}/{\tau}$ \cite{Grathwohl1998,Hoeche2014}, in which $\epsilon$ and $\tau$ are the porosity and tortuosity of the protective film, respectively. The effective diffusion coefficient can be then calculated by interpolating the two aforementioned values:
\begin{equation} \label{eq:diff_coeff}
D_{i}^{e}=D_{i}\left(\left(1-\beta \frac{C_{\mathrm{Film}}}{[\mathrm{Film}]_{\max }}\right)+\beta \frac{C_{\mathrm{Film}}}{[\mathrm{Film}]_{\max }} \frac{\epsilon}{\tau}\right).
\end{equation}
The $\beta$ coefficient is called momentum here and controls the effect of the saturation term $(1-\frac{C_{\mathrm{Film}}}{[\mathrm{Film}]_{\max }})$. The derivation of these equations is discussed in detail in our previous work \cite{Barzegari2020arXiv}.

\subsection{Interface movement formulation}

In order to take advantage of the level set method for tracking the corrosion front, the velocity of the interface at each point should be determined. Then, by solving Eq. \ref{eq:lsm_simplified}, the interface is obtained at the points with a zero value of the $\phi$ function. The interface velocity in mass transfer problems can be calculated using the Rankine–Hugoniot equation \cite{Scheiner2007}, and by considering the transportation of $\mathrm{Mg}^{2+}$ ions, it can be written as:
\begin{equation} \label{eq:rankine}
\left\{\mathbf{J}(x, t)-\left([\mathrm{Mg}]_{\mathrm{sol}}-[\mathrm{Mg}]_{\mathrm{sat}}\right) \mathrm{V}(x, t)\right\} \cdot n=0
\end{equation}
where $\mathbf{J}$ is the mass flux at the interface. Rearranging Eq. \ref{eq:rankine} and inserting the value of the normal interface velocity into Eq. \ref{eq:lsm_simplified} yields:
\begin{equation} \label{eq:lsm_final}
\frac{\partial \phi}{\partial t}-\frac{D_{\mathrm{Mg}}^{e} \nabla_{n} C_\mathrm{Mg}}{[\mathrm{Mg}]_{\mathrm{sol}}-[\mathrm{Mg}]_{\mathrm{sat}}}|\nabla \phi|=0,
\end{equation}
which is the final form of the level set equation to be solved. In the case of simulations with a low diffusion rate, the interface moves slowly in the beginning, which results in a linear degradation, whereas based on the experimental results, the degradation rate is fast at the beginning and slows down eventually \cite{Mei2019}. So, to mimic the same behavior in the low diffusion regimes, we took advantage of the theoretical Stefan formulation (Eqs. \ref{eq:stefan} and \ref{eq:stefan_alpha}) to push the interface in the first time step. According to Eq. \ref{eq:stefan}, the velocity of the interface can be calculated as $({2\alpha}/{\sqrt{t}})$, but as we are dealing with a 3D model and not a 1D one, we pick a fraction (denoted by $\gamma$) of this ideal value to be used as the driving force of the interface at the beginning of the simulation. So, the normal velocity of the interface can be written in the general form as:
\begin{equation} \label{eq:normal_vel}
\mathrm{V}^\mathrm{N}(x, t)=\left\{\begin{array}{ll}
\gamma\frac{2\alpha}{\sqrt{t}} & t=0 \\
\frac{D_{\mathrm{Mg}}^{e} \nabla_{n} C_\mathrm{Mg}}{[\mathrm{Mg}]_{\mathrm{sol}}-[\mathrm{Mg}]_{\mathrm{sat}}} & t > 0
\end{array}\right.
\end{equation}
in which the $\alpha$ value should be calculated from Eq. \ref{eq:stefan_alpha}. By selecting $\gamma$ equal to zero, the Stefan formulation can be eliminated, and a value of 1 for $\gamma$ restores the ideal 1D velocity definition.

\subsection{Boundary conditions}

The implementation of boundary conditions is relatively challenging and  complex for the developed model as they should be imposed inside the domain of interest on virtual interfaces defined by mathematical expressions (i.e. on the moving interface defined by the zero iso-contour of the level set equation). The penalty method was  used to overcome this issue and define the desired boundary conditions on the moving corrosion front. 

\begin{figure}[h]
\center \includegraphics[width=10cm]{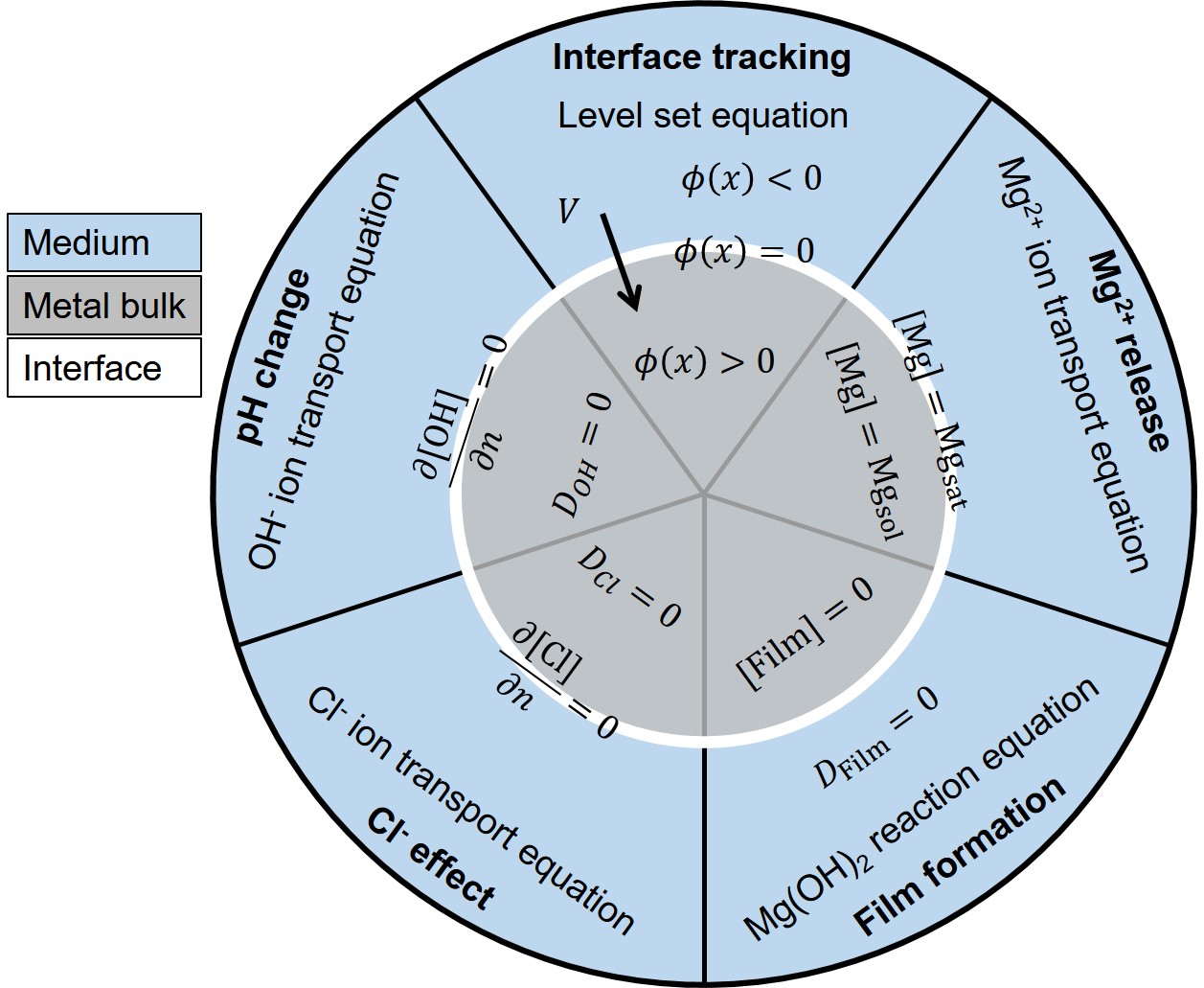}
\caption{A schematic overview of the exposed boundary conditions and constraints required for the simulation of each equation of the developed mathematical model for Mg biodegradation.} \label{fig:bc}
\end{figure}

Fig. \ref{fig:bc} demonstrates a schematic presentation of the boundary conditions and general considerations of each PDE of the biodegradation mathematical model. This figure is divided into 5 different parts, presenting the 5 PDEs of the model. The Mg block is depicted in the center, and the interface separates it from the surrounding medium. There is no specific boundary condition for the level set and film formation equations, but in comparison to the other 3 transport equations, it should be noted that diffusivity is not considered for $\mathrm{Mg}(\mathrm{OH})_{2}$, which is also reflected in Eq. \ref{eq:pde_film}. The level set function $\phi$ is defined in a way that is positive inside and negative outside the solid region. For the $\mathrm{Mg}^{2+}$ ions transport equation, a Dirichlet boundary condition is applied on the mathematical interface to make the concentration equal to the saturation concentration of $\mathrm{Mg}^{2+}$ ions, a value that was already used in Eq. \ref{eq:lsm_final}. For the $\mathrm{Cl}^{-}$ and $\mathrm{OH}^{-}$ ions transport equations, a no-flux boundary condition is applied to the interface by making the diffusion coefficient equal to zero inside the Mg block, preventing ions to diffuse inside the solid material.

\subsection{Implementation}

To simulate the developed mathematical model, which is comprised of Eqs. \ref{eq:pde_mg}, \ref{eq:pde_film}, \ref{eq:pde_cl}, \ref{eq:pde_oh}, and \ref{eq:lsm_final}, a combination of finite difference and finite element methods was used, leading to discrete forms of these equations, which were subsequently solved using appropriate linear solvers.

To discretize the temporal terms of the aforementioned parabolic PDEs, a first-order backward Euler finite difference scheme was used, whereas the spatial terms were converted to a weak form using a standard first-order finite element scheme. Then, the open-source PDE solver FreeFEM \cite{Hecht2012} was used to implement the weak form and obtain a linear system of equations for each PDE. The obtained linear systems were solved in parallel using the HYPRE preconditioner \cite{Falgout2002} and the GMRES solver \cite{Saad1986} via the open-source high-performance computing (HPC) toolkit PETSc \cite{petsc}. Additionally, to increase the efficiency of the computation and decrease the simulation execution time, the computational mesh was decomposed and distributed among available computing resources using the interface of HPDDM package in FreeFEM \cite{Jolivet2013}. The details of this implementation are presented in our previous work \cite{Barzegari2020arXiv} as well as in the supplementary materials of this paper. A simple iterative solver based on the Newton method was also developed to solve Eq. \ref{eq:stefan_alpha} to obtain the value of $\alpha$ parameter if it was required in the simulations.

The computational mesh was generated using a set of first-order tetrahedral elements and was adaptively refined on the metal-solution interface to increase the numerical accuracy of the simulation of the level set equation (Eq. \ref{eq:lsm_final}). The Netgen mesh engine \cite{Schoeberl1997} in the SALOME platform \cite{Ribes2007} was used to generate the mesh.

Similar to the technique employed by Bajger et al. \cite{Bajger2016}, the gradient of concentration of $\mathrm{Mg}^{2+}$ in Eq. \ref{eq:lsm_final} was calculated at a distance $h$ in the normal direction from the interface, with $h$ being the smallest element size of the mesh:
\begin{equation}
\nabla_{n} C = \frac{C\left( \mathbf{x} + h.n \right) - C\left( \mathbf{x} + 2h.n \right)}{h} \quad \mathbf{x} \in \Omega \subset \mathbb{R}^{3}.
\end{equation}
Considering the adaptively refined mesh, the $h$ value is very small, so the gradient is computed at the regions close enough to the interface. In addition to this technique, the mass lumping feature of FreeFEM was used to prevent the oscillation of concentration values on the diffusive metal-medium interface.

\subsection{Experimental setup}

The degradation rate of CP Mg was evaluated based on the hydrogen evolution tests  performed either in  NaCl or SBF solutions with eudiometers. The composition of the electrolytes is  shown in the following table (Table \ref{tab:composition}). 0.5 g metallic chips (with a surface area of $47.7\pm5.0 \mathrm{cm}^2/\mathrm{g}$ and chip thickness ca. 200 microns) of CP Mg were put in 500 ml electrolyte for 22-24 hours for monitoring the amount of evolved hydrogen. The bulk pH of electrolytes before and after corrosion was measured by a pH meter (Metrohm-691, Switzerland). Local pH was measured by positioning pH microprobes (Unisense, Denmark, pH-sensitive tip size 10x50 micron) 50 micron above the surface of Mg and monitoring the pH values either in one spot or by horizontal or vertical line-scans or mapping by following a horizontal grid. The measurements were performed at room temperature of $22\pm2^{\circ}\mathrm{C}$ maintained by the laboratory climate control system. More detailed information about experimental set up and procedures  can be found  elsewhere \cite{Mei2019,Mei2019a}.

\begin{table}[h] 
\caption{Chemical composition of NaCl and SBF electrolytes used to perform hydrogen evolution tests, weight loss, local and bulk pH measurements.}
\centering
\begin{tabular}{ccc}
& \multicolumn{2}{c}{Concentration/ mM} \\
\cline{2-3}
 & 0.85 wt. \% NaCl & SBF \\ \hline
$\mathrm{Na}^{+}$ & 145.4 & 142.0 \\
$\mathrm{K}^{+}$ & - & 5.0 \\
$\mathrm{Mg}^{2+}$ & - & 1.5 \\
$\mathrm{Ca}^{2+}$ & - & 2.5 \\
$\mathrm{Cl}^{-}$ & 145.4 & 147.8 \\
$\mathrm{HCO}^{-}_3$ & - & 4.2 \\ 
$\mathrm{HPO}^{2-}_4/\mathrm{H}_2\mathrm{PO}^{-}_4$ & - & 1.0 \\
$\mathrm{SO}^{2-}_4$ & - & 4.2 \\ 
\begin{tabular}[c]{@{}c@{}}Synthetic pH buffer \\ (i.e. Tris/HCl, HEPES)\end{tabular}& No & No  \\
Initial pH value & 5.6-5.9 & 7.35-7.45 \\
\hline
\end{tabular}
\label{tab:composition}
\end{table}

\subsection{Parameter estimation}

The constructed mathematical model contains some parameters that need to be calibrated prior to final validation of the model: diffusion coefficient of $\mathrm{Mg}^{2+}$ and $\mathrm{Cl}^{-}$ ions ($D_{\mathrm{Mg}}$ and $D_{\mathrm{Cl}}$ to be inserted into Eq. \ref{eq:diff_coeff} to get effective diffusion coefficients), the reaction rates of Eqs. \ref{eq:oxidation_react} and \ref{eq:break_react} ($k_1$ and $k_2$), the momentum parameter, $\beta$, for controlling the saturation term behavior (in Eqs. \ref{eq:pde_mg}, \ref{eq:pde_film}, and \ref{eq:diff_coeff}), and the $\gamma$ parameter for the initial interface velocity (Eq. \ref{eq:normal_vel}). An inverse problem setup was required to estimate the proper value of these parameters. 

Performing a parameter estimation requires running the computational models several times. Considering the computationally-intensive model of the current study, a sensitivity analysis was performed prior to the parameter estimation to exclude non-essential parameters and reduce the time required to complete the inverse problem run. This sensitivity analysis was accomplished separately in the low diffusion (similar to the SBF solution) and high diffusion (similar to NaCl solution) regimes. 

After determining the essential parameters to include, a Bayesian optimization approach \cite{Mockus1989} was used to construct the inverse problem and calibrate the parameters. The reason for choosing a Bayesian approach was to minimize the number of optimization iterations, in each of which the simulation should run once. The Bayesian optimization is a more efficient option for such computational intensive cases in comparison to gradient-based or fully-stochastic methods as it takes into account all the preceding iterations in a probability tree \cite{Mehrian2017}.

The objective function of the optimization problem was the difference between the predicted and experimentally obtained values of evolved hydrogen. In the computational model, the evolved hydrogen can be computed directly at any time through the mass loss as each mole of corroded Mg is correlated to one mole of released hydrogen (Eq. \ref{eq:oxidation_react}). The mass loss can be obtained using the following volume integral:
\begin{equation} \label{eq:mass_loss}
\mathrm{Mg}_{\mathrm{lost}}=\int_{\Omega_{+}(t)} [\mathrm{Mg}]_{\mathrm{sol}} \mathrm{d} V-\int_{\Omega_{+}(0)} [\mathrm{Mg}]_{\mathrm{sol}} \mathrm{d} V,
\end{equation}
where $\Omega_{+}(t)=\{\mathbf{x}: \phi(\mathbf{x}, t) \geq 0\}$, and then, the amount of produced hydrogen is calculated using the ideal gas law:
\begin{equation} \label{eq:evolv_hydr}
H_{f}=\frac{\mathrm{Mg}_{\mathrm{lost}}}{\mathrm{Mg}_{\mathrm{mol}}} \frac{R T}{P}
\end{equation}
in which $R$, $P$, $T$, $\mathrm{Mg}_{\mathrm{mol}}$ are the universal gas constant, the pressure, the medium temperature, and  the molar mass of Mg, respectively. 

\subsection{Simulation setup}

\begin{figure}[h]
\center \includegraphics[width=13cm]{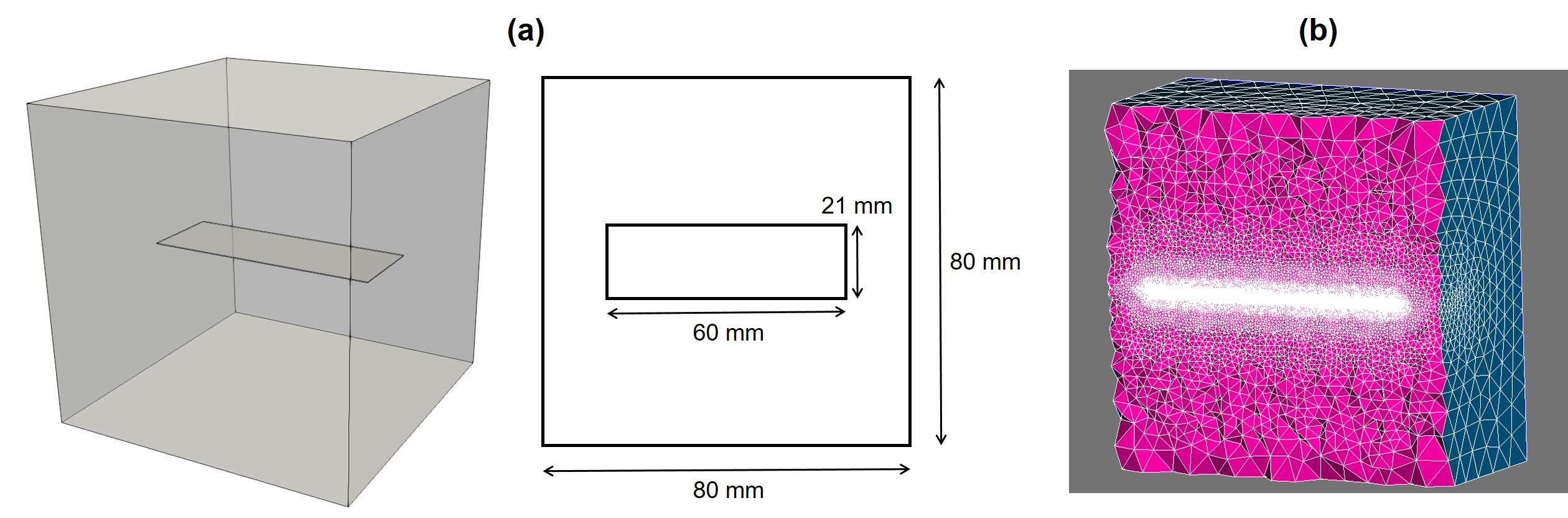}
\caption{Computational representation of the experimental set-up, used to perform parameter estimation and numerical validation of the developed model. a) A cuboid of Mg ($60 \mathrm{mm} \times 21 \mathrm{mm} \times 0.2 \mathrm{mm}$) inside a solution, b) a cross-section of the computational mesh, refined on the corrosion front to increase the required level set accuracy.} \label{fig:simulation_setup}
\end{figure}

In order to simulate the developed mathematical model, the experimental setup was reconstructed \textit{in silico} with some minor differences. As there is no perfusion in the solution chamber, the mixing effect was neglected, so, as can also be seen in the mathematical model, the advection terms were not considered. Furthermore, the experiments were conducted using small metallic chips, yet, as the biodegradation behavior heavily depends on the exposed surfaces, we represented these chips by a cuboid with the same surface-to-mass ratio. By considering the approximate surface-to-mass of $50 \mathrm{cm}^2/\mathrm{g}$ and the total mass of $0.5 \mathrm{g}$, the chips were replaced by a cuboid with the size of $60\mathrm{mm} \times 21\mathrm{mm} \times 0.2 \mathrm{mm}$, which approximately has the same ratio, surface area, volume, and mass. Also, the solution chamber with a capacity of $500 \mathrm{ml}$ was represented by a cubic container with an edge size of $80 \mathrm{mm}$. Fig. \ref{fig:simulation_setup} depicts the constructed geometry as well as the computational mesh generated to represent it. The mesh is refined on the interface and contains $\num{18049471}$ elements, resulting in $\num{3077227}$ degrees of freedom (DOF) for each PDE. 

Simulations were carried out on the VSC (Flemish Supercomputer Center) supercomputer. Taking advantage of HPC techniques to parallelize the simulation is an inevitable aspect of such a computational-intensive model, so based on what described in the implementation section, the mesh was decomposed among 170 computing cores, i.e. $\num{24137}$ DOF per core (which includes the ghost nodes to satisfy the boundary condition in each sub-mesh). On the VSC supercomputer, we made use of 5 nodes, 36 cores each, each node holding CPUs with a clock speed of 2.6 GHz, with 960 GB of the total available memory. 

The $\mathrm{OH}^{-}$ transport equation (Eq. \ref{eq:pde_oh}) was not solved during the parameter calibration process. Afterwards, two full simulations (for the NaCl and SBF solutions) were conducted to calculate the pH changes based on the change of the concentration of $\mathrm{OH}^{-}$ ions in the medium. This acted as the validation of the numerical model because no calibration was performed on the output of this equation. The pH was calculated using Eq. \ref{eq:ph}, based on the solution of Eq. \ref{eq:pde_oh} and a reported value of $7.00e\times10^{-5} \mathrm{cm}^2/\mathrm{s}$ ($25.2 \mathrm{mm}^2/\mathrm{hour}$) for the diffusion coefficient of $\mathrm{OH}^{-}$ ions ($D_{\mathrm{OH}}$ to be used in Eq. \ref{eq:diff_coeff}) in aqueous solutions \cite{Lee2011}. 

According to the experimental setup, the initial concentration of the $\mathrm{Mg}^{2+}$, $\mathrm{Cl}^{-}$, and $\mathrm{OH}^{-}$ ions were set to 0 (no $\mathrm{Mg}^{2+}$ ions at the beginning), $146 \mathrm{mM}$ ($5.175\times10^{-6} \mathrm{g}/\mathrm{mm}^3$), and $1\times10^{-7} \mathrm{g}/\mathrm{mm}^3$, respectively. The porosity ($\epsilon$) and tortuosity ($\tau$) of the protective film were considered to be 0.55 and 1, respectively \cite{Sun2012}. The saturation concentration $[\mathrm{Mg}]_{\mathrm{sat}}$ was set to the solubility of magnesium chloride in water, which is $134\times10^{-6}\mathrm{g}/\mathrm{mm}^3$ at $25^{\circ}\mathrm{C}$ \cite{GrahamC.Hill2001}. The density of Mg ($[\mathrm{Mg}]_{\mathrm{sol}}$) and $\mathrm{Mg(OH)}_2$ were set to $1735\times10^{-6}\mathrm{g}/\mathrm{mm}^3$ and $2344\times10^{-6}\mathrm{g}/\mathrm{mm}^3$, respectively \cite{Bajger2016}. A time step convergence study was performed to determine the implicit time step size. Based on the results, a time step with a size of $0.025$ hours was chosen. The overall simulated time is 22 hours in accordance with the experimental design of performed immersion tests.

\subsection{Case study}

To further investigate the predictions of the current model on more complex shapes, the biodegradation of a simple screw was studied in the SBF solution using the parameters obtained for the low diffusion regimes. Similar to the simulation of Mg cuboid, the mesh was refined on the metal-medium interface, and it consisted of $\num{1440439}$ elements with $\num{246580}$ DOFs for each PDE. All the simulation parameters and materials properties were identical to the simulation of biodegradation in the SBF solution, and the target was to simulate $42$ days ($1008$ hours) of the process. This was selected as a sufficiently long simulated time to observe the effects of biodegradation on larger time scales.

\section{Results}

\subsection{Optimization results}

Based on the performed sensitivity analysis, two parameter sets were obtained for the high diffusion (in NaCl solution) and low diffusion (in SBF solution) simulations, respectively. These parameters are listed in Table \ref{tab:sensitivity}. According to the results, the reaction rate of  Eq. \ref{eq:oxidation_react} ($k_1$), which demonstrates the rate of oxidation-reduction, has less contribution to the process in comparison to the rate of the weakening of the protective film ($k_2$). Because of this, the parameter $k_1$ was not selected for the parameter estimation. Also, the model was sensitive to the effect of parameter $k_2$ in different ranges of values and not on a specific point, and as a result, three constant values were chosen as the delegates of these ranges in the optimization process. The model was not sensitive to the diffusion rate of $\mathrm{Cl}^{-}$ ions, which was also expected because although $\mathrm{Cl}^{-}$ has an important role in the weakening of the partially protective MgO film, its transport equation (Eq. \ref{eq:pde_cl}) is purely diffusive and does not include any reaction term.

\begin{table}[t] 
\caption{The effective parameters as the result of the sensitivity analysis and their corresponding range to be considered in the Bayesian optimization for parameter calibration}
\centering
\begin{tabular}{lll}
\cline{2-3}
                                                                                 & Parameter & Optimization range \\ \hline
\multirow{4}{*}{\begin{tabular}[c]{@{}l@{}}Low diffusion\\ (SBF solution)\end{tabular}}   & $D_\mathrm{Mg}$       & $[0.0001, 0.01]$     \\
                                                                                 & $k_2$        & $10^{10}, 10^{15}, 10^{20}$     \\
                                                                                 & $\beta$     & $[0.1,10]$     \\
                                                                                 & $\gamma$     & $[0,1]$     \\ \hline
\multirow{3}{*}{\begin{tabular}[c]{@{}l@{}}High diffusion\\ (NaCl solution)\end{tabular}} & $D_\mathrm{Mg}$       & $[0.003, 0.1]$     \\
                                                                                 & $k_2$        & $10^{10}, 10^{15}, 10^{20}$     \\
                                                                                 & $\beta$     & $[0.1,10]$     \\ \hline
\end{tabular}
\label{tab:sensitivity}
\end{table}

The parameter optimization process was performed on the specified range of selected parameters, while the rest of the parameter values were obtained from the literature \cite{Bajger2016,Lee2011}. Table \ref{tab:parameters} shows the output of this process, which was used to simulate the full model. For two estimation processes, 120 optimization iterations were taken cumulatively, which took 276 hours of simulation execution time using 170 computing nodes for each simulation.

\begin{table}[t]
\caption{Values used to evaluate the model performance, obtained from the output of the optimization process and the literature.}
\centering
\begin{tabular}{lccccccc}
Parameter & $D_\mathrm{Mg}$ & $D_\mathrm{Cl}$ & $D_\mathrm{OH}$ & $k_1$ & $k_2$ & $\beta$ & $\gamma$ \\ \hline
Unit      & $\frac{\mathrm{mm}^2}{\mathrm{hour}}$ & $\frac{\mathrm{mm}^2}{\mathrm{hour}}$ & $\frac{\mathrm{mm}^2}{\mathrm{hour}}$ & $\frac{1}{\mathrm{hour}}$ & $\frac{\mathrm{mm}^6}{\mathrm{g}^2 \mathrm{hour}}$ & - & - \\ \hline
SBF solution       & $0.000338$ & $0.05$ & $25.2$ & $7$ & $10^{15}$ & $0.125$ & $0.65$  \\
NaCl solution      & $0.06273$ & $0.05$ & $25.2$ & $7$ & $10^{20}$ & $0.2$ & $0$
\end{tabular}
\label{tab:parameters}
\end{table}

\subsection{Degradation prediction}

\begin{figure}[t]
\center \includegraphics[width=12cm]{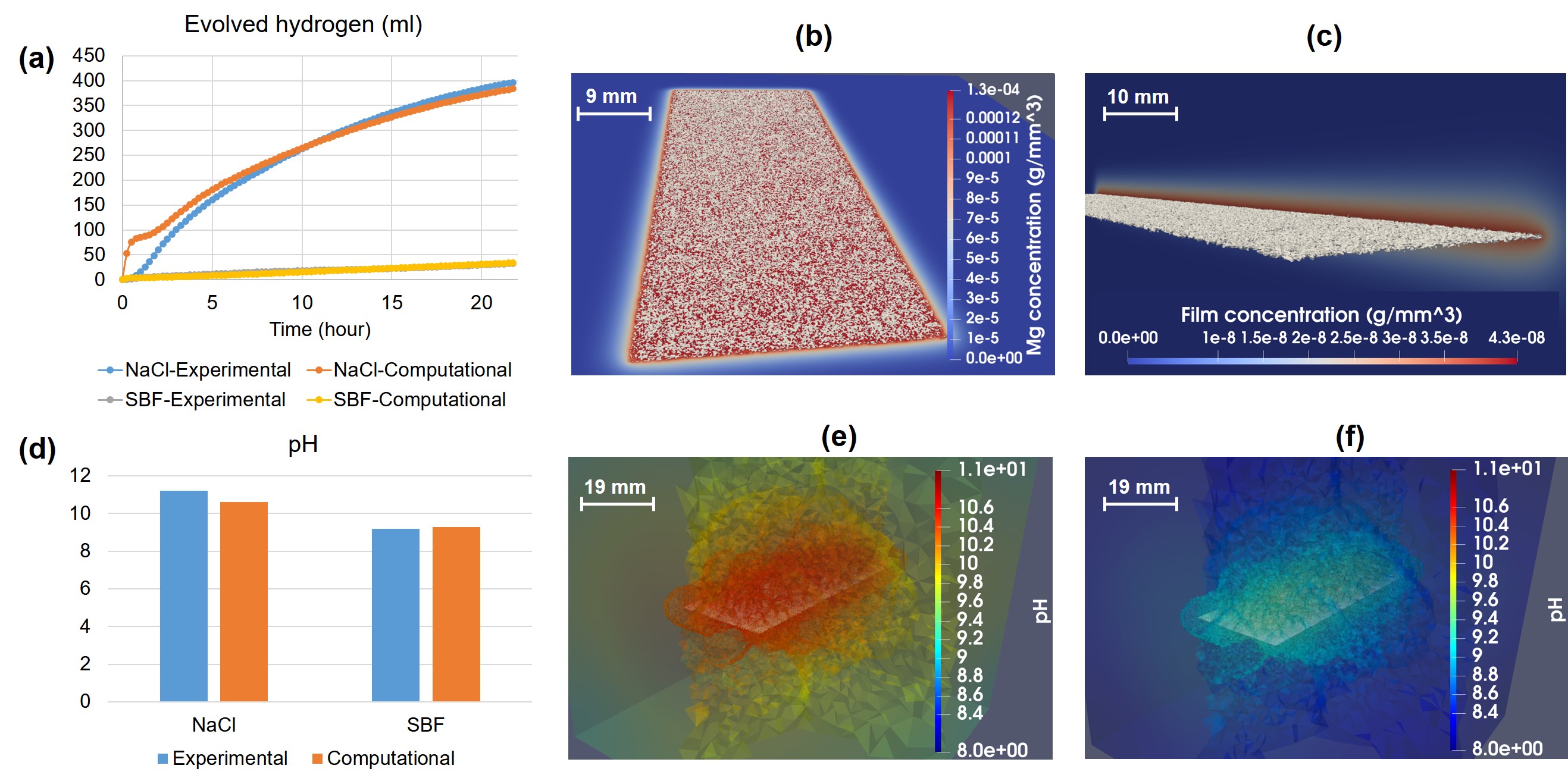}
\caption{Comparing the quantitative output of the model for the rate of degradation and the pH changes in NaCl and SBF solutions with experimentally measured values as well as the simulation results for ion release, mass loss, protective film formation, and pH changes after 22 hours of simulated time: a) calibrated output of the formed hydrogen gas during the degradation (the SBF curves are overlapped), b) the simulation results of $\mathrm{Mg}^{2+}$ ions release, c) the simulation results of protective film concentration at the end of the simulation (the color contour shows the concentration of species, and the gray surface is the zero iso-contour of the level set function, which indicates the surface of the Mg block), d) de novo prediction of the global pH changes in the medium, showing a good agreement between the model output and the experimental results, e) pH changes in different regions of the medium in NaCl solution, f) pH changes in SBF solution.} \label{fig:results}
\end{figure}

Fig. \ref{fig:results} shows the model output for the predicted produced hydrogen, protective film formation, and the pH changes. The graph of the evolved hydrogen is used as input during the parameter optimization process, but the pH results are produced by the simulations using the optimized parameters to demonstrate the validation of the developed mathematical and computational models. The predicted pH result (Fig. \ref{fig:results}-d) shows a difference of $5.35 \%$ for the simulation in NaCl and $1.03 \%$ for SBF simulation. Each simulation took about 3 hours to complete.

In Fig. \ref{fig:results}, a post-processed view of the final shape of the Mg cuboid in the NaCl solution is presented, in which the degraded geometry is plotted on the $\mathrm{Mg}^{2+}$ ions (Fig. \ref{fig:results}-b) and protective film concentration (Fig. \ref{fig:results}-c) contours. A transparent contour of the pH values in the solution is depicted for both the NaCl (Fig. \ref{fig:results}-e) and SBF (Fig. \ref{fig:results}-f) solutions. The range of colors is kept equal for both contours to make it easy to compare the change of pH in both solutions.

\begin{figure}[t]
\center \includegraphics[width=12cm]{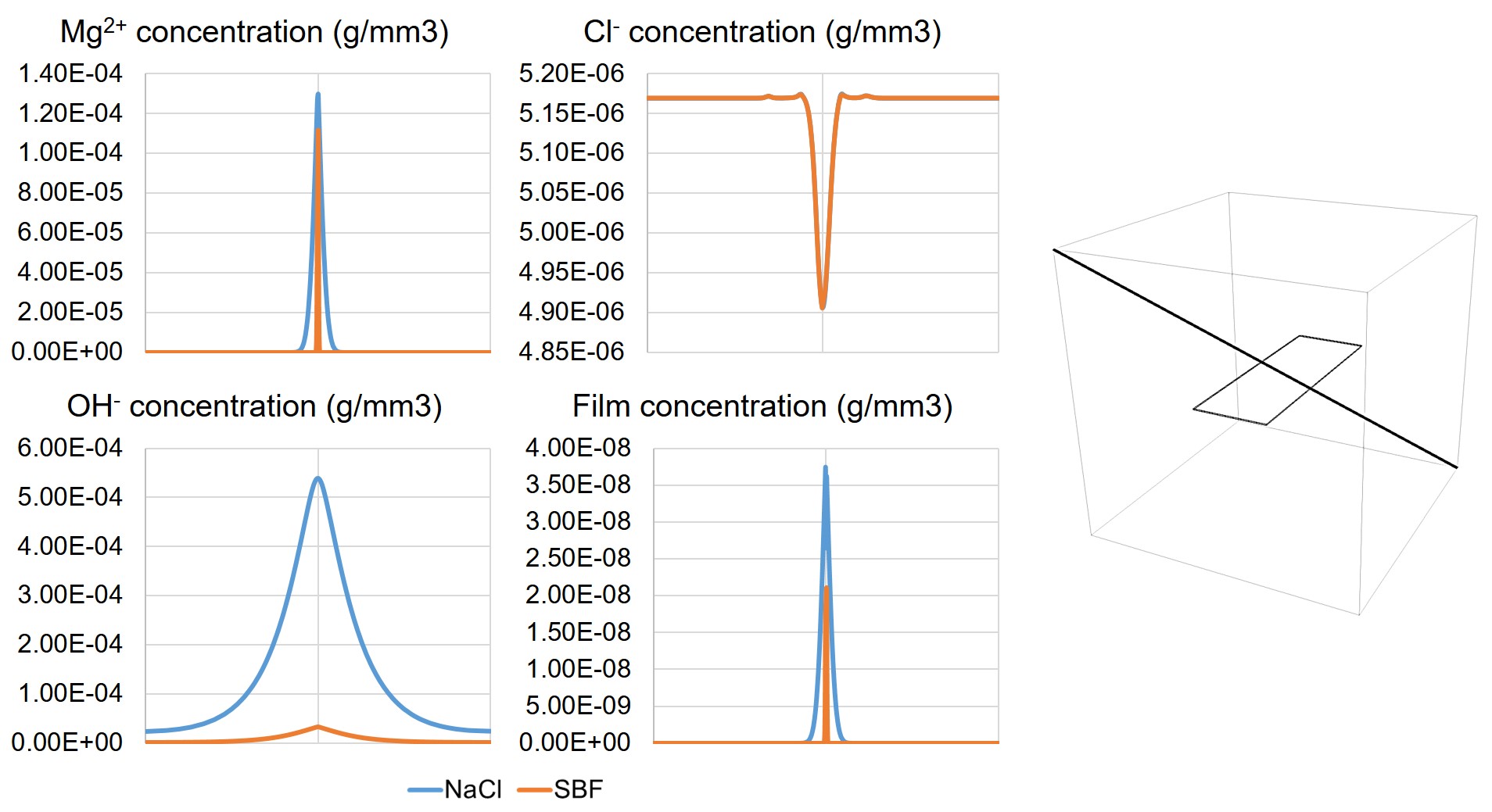}
\caption{The change of concentration for the involved chemical components, $\mathrm{Mg}^{2+}$, $\mathrm{Cl}^{-}$, $\mathrm{OH}^{-}$ ions, and the protective precipitation structure (which can be correlated to the thickness of the layer) plotted over a diagonal line as shown in the right.} \label{fig:results_diagonal}
\end{figure}

The concentration values of the state variables of the derived transport PDEs ($\mathrm{Mg}^{2+}$, $\mathrm{Cl}^{-}$, $\mathrm{OH}^{-}$, and $\mathrm{Mg(OH)}_2$) are plotted along a diagonal line in the solution container in Fig. \ref{fig:results_diagonal}, showing how they change in the zones close to the corrosion surface and far from it. The animated output of the degradation of the Mg cuboid is presented as a set of movie files in the supplementary materials of this paper (post-processed using NVIDIA IndeX software on a GPU).

\subsection{Example application}

\begin{figure}[t]
\center \includegraphics[width=12cm]{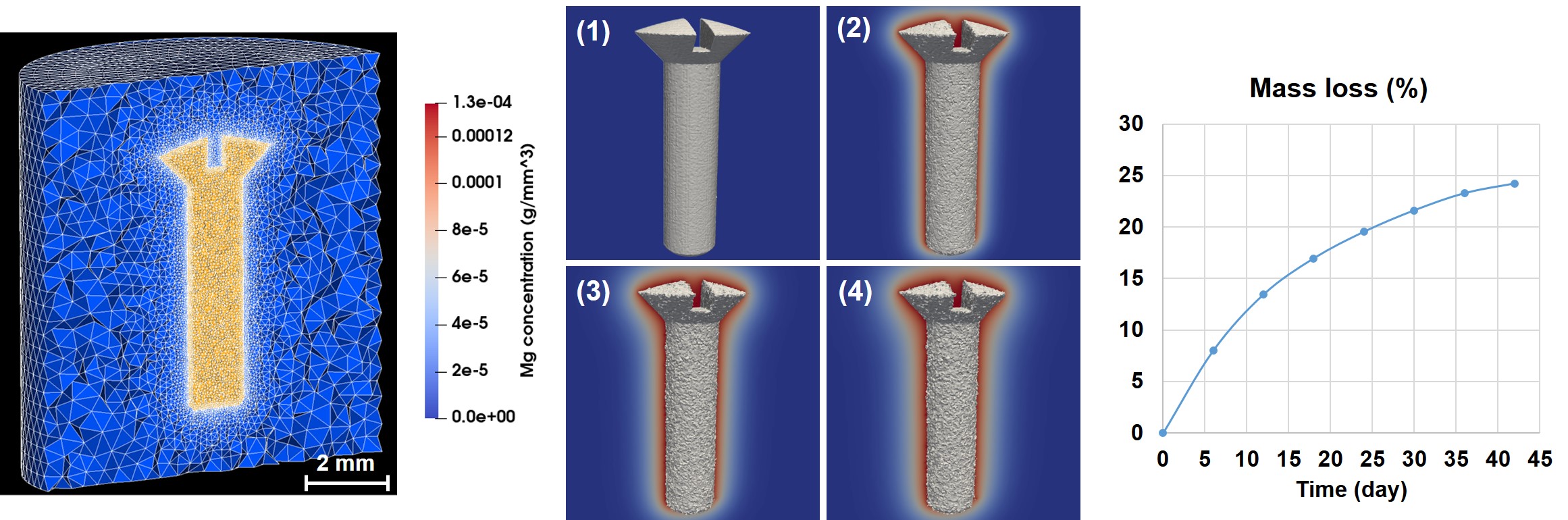}
\caption{A cross-section of the computational mesh and simulation results of the degradation process of the use-case screw in SBF solution as well as the mass loss graph over time. The contours display the concentration of $\mathrm{Mg}^{2+}$ ions on a cross-section view of the medium beside the moving surface of the screw at 1) 1st day (initial state), 2) 6th day, 3) 12th day, and 4) 18th day.} \label{fig:results_screw}
\end{figure}

The simulation of 42 days ($\num{19200}$ time steps) of the degradation of the simple screw took 9 hours to run using 170 computing cores. Fig. \ref{fig:results_screw} depicts the post-processed interface and $\mathrm{Mg}^{2+}$ ions release (similar to Fig. \ref{fig:results}-b) as well as the mass loss during the degradation of the screw in the SBF solution.

\section{Discussion}

In this study, a physicochemical model of the biodegradation process of commercially-pure Mg was developed by constructing a mathematical model formulating the mass transfer phenomena as well as tracking the location of the surface of the implant during degradation. For the mass transfer model, the equations were derived from the chemistry of biodegradation of the Mg in saline (NaCl) and buffered (SBF) solutions, which includes the oxidation of the metallic part, reduction of water, changes in pH, and formation of a protective film on the surface of the scaffold which contributes to a slower rate of degradation. Beside these aspects, it was also crucial to consider the effect of different ions in the medium on the rate of degradation. Additionally, investigating the structural changes of the scaffolds and implants in practical applications, like resorption of temporary fixation devices, requires tracking the movement of the corrosion surface. This was done by constructing an equation based on the level set principle, which captured the movement of the medium-metal interface by defining an implicit surface. The derived equations were coupled and solved using a combination of finite difference and finite element methods. The degradation data to validate the models was collected from immersion tests of small Mg chips, reconstructed as a single cuboid in the computational study with a similar surface over volume properties. The model parameters were calibrated using a Bayesian optimization algorithm, and the obtained parameters were used to simulate the pH changes in NaCl and SBF solutions. 

The developed model falls in the categories of physical models of the corrosion process, which provide more insights of the process in comparison to the phenomenological models. The reason is that the phenomenological models focus on the elimination of elements to capture the loss of materials, which makes it impossible to model the formation of new chemical compounds or interaction of species \cite{Abdalla2020}. The physical models, like the one developed in this study, are capable of capturing the underlying chemical interactions. By doing this, processes like the effect of coating, the formation of a protective layer, and pH changes can be modeled. Adding an appropriate interface tracking method enables the physical models to act like the phenomenological models in capturing the corrosion interface movement. In the current study, this has been accomplished using a level set function. Technically speaking, this approach has certain benefits over the  ALE method,  which is the method used by several similar studies, including Grogan et al. \cite{Grogan2014}. In comparison to the ALE method, the level set function tracks the interface instead of a Lagrangian mesh, and elements can freely be marked as solid or liquid. Additionally, employing the ALE method for degradation simulation requires remeshing the geometry as the interface moves, which is not efficient for 3D models and is limited to the available features of the employed numerical solver.  

One of the challenging aspects of validating physical models is getting the correct value for the parameters of said models, requiring dedicated experimental input. To overcome this challenge, an efficient inverse problem was constructed based on the Bayesian optimization approach to estimate the unknown parameters. 
To save time and resources, the parameter estimation process was performed on the most effective parameters, which were selected based on a sensitivity analysis. This selection process implied the importance of parameters in high and low diffusion rates.

The degradation rate is fast at the beginning, but then it slows down due to the formation of a partially protective film and also because of the saturation concentration. This phenomenon is well captured by the model at high diffusion rates, but in low diffusion rates (in SBF solution), this effect can be reproduced by pushing the corrosion front according to the Stefan formulation of the moving interface problems. This was controlled by the parameter $\gamma$ (Eq. \ref{eq:normal_vel}).
It should be noted that the inclusion of the $\gamma$ parameter is crucial for short-term simulations only, helping the model mimic the chemical behavior correctly.
Defining and considering  $\gamma$ is necessary because from the computational costs perspective, performing the parameter calibration on simulations with thousands of time steps requires a lot more resources and time.

The degradation of the CP Mg was assumed to be mostly diffusion-based. As a result, the value of $D_\mathrm{Mg}$ plays an important role in the behavior of the model. Although it was possible to get the diffusion coefficient of $\mathrm{Mg}^{2+}$ from the previously conducted experiments in the literature (similar to what was done for $D_\mathrm{OH}$), we decided to not do so because of two reasons: 1) the values reported in the literature are mostly valid for saline solutions only, and 2) the reported values were not in a good agreement with one another \cite{Grogan2014,Sun2012}. Thus, the diffusion coefficient was obtained using the parameter estimation process for both the NaCl and SBF solutions. The obtained value of $D_\mathrm{Mg}$ ($0.06273 \mathrm{mm}^2/\mathrm{hour}$) was in line with the values that Grogan et al. have already suggested ($0.010575-0.50575 \mathrm{mm}^2/\mathrm{hour}$) \cite{Grogan2014}, showing that the constructed inverse problem was successful in reproducing previous results of similar studies. The obtained value of $D_\mathrm{Mg}$ in the Bajger et al. work \cite{Bajger2016} is $0.00066\mathrm{mm}^2/\mathrm{hour}$, which is mostly related to the simplicity of the employed parameter estimation method as well as having a 2D model instead of a 3D one.

In the \textit{in vitro} biodegradation of Mg-based biomaterials, the local pH of the surrounding solution increases less than that in NaCl solution. This is because the $\mathrm{Mg(OH)}_2$ formed in NaCl stabilizes pH at 10.4 \cite{Santucci2018}, while Mg-Ca-P-C containing products stabilize the pH at 7.8-8.5 since OH- is consumed for the formation of this product \cite{Lamaka2018,Mei2019}. This phenomenon was captured in Eqs. \ref{eq:pde_oh} and \ref{eq:ph} to calculate pH based on the concentration of $\mathrm{OH}^{-}$ ions, showing the local pH changes at any location (Fig. \ref{fig:results}-e,f). In the current study, the global pH is considered as the validation criterion, which means that the average value of the solution pH is calculated using a volume integral and is compared with the ones obtained from the experiments. Fig. \ref{fig:results}-d shows that such a prediction has a good agreement with the experimental data. It is worth noting that the model can be extended to non-pure Mg by considering the effect of alloying elements on the reaction rates as well as adding more terms to the transport equations to capture the electrochemical potential changes, converting the PDEs into the Nernst–Planck equation \cite{Deshpande2011}. By doing so, more complex forms of the corrosion process, such as galvanic corrosion, can be predicted by the model.

One of the biggest simplifications of the current study was made by ignoring the contribution of pH changes to the biodegradation mechanism of Mg. Although doing that is relatively simple and straightforward in the approach taken by this study, it results in non-linear terms in the derived PDEs. This non-linearity inserts another level of complexity to the computational model as the order of the state variables are in the range of $10^{-6}$ to $10^{-10}$, which makes it difficult to yield convergence in the iterative non-linear solvers. By developing a robust non-linear solver, this effect can be added simply by including more relevant terms as the effect of Eq. \ref{eq:pde_oh} into Eq. \ref{eq:pde_mg}. 

Additionally, buffered solutions and the physiological fluids inside the human and animal bodies contain more ions interacting with more complex chemistry \cite{Mei2020}. In this study, this effect was encapsulated in a limited number of parameters (such as $k_1$ and $k_2$ in Eqs. \ref{eq:pde_mg}, \ref{eq:pde_film}, and \ref{eq:pde_oh}), but while the results show its success to reproduce experimental observations, it still needs additional elaboration to be able to capture more chemical interactions. For SBF solutions, the effect of presented inorganic ions such as $\mathrm{HCO}^{-}_3$, $\mathrm{HPO}^{2-}_4/\mathrm{H}_2\mathrm{PO}^{-}_4$, and $\mathrm{Ca}^{2+}$ can be added similar to the way the effect of $\mathrm{Cl}^{-}$ was considered. Additionally, formulating the effect of $\mathrm{HPO}^{2-}_4$ that exists in the physiological environments will make the model capable of making more accurate predictions for \textit{in vivo} studies.

Although the pH simulations are not enough experimental input to call the model fully validated, the obtained validation results show that the derived mathematical model and the corresponding parallelized computational model give a correct \textit{in silico} representation of the studied process. The performed predictive simulations, including the case study, demonstrate the potential of the developed computational model and software to study the biodegradation behavior of implants. This can be further combined with other computational models to provide a multidisciplinary environment to investigate the mechanical integrity of implants or induced neotissue growth for different applications in orthopedics and tissue engineering. 

\section{Conclusions}

The use of biodegradable metals for designing medical devices and implants has the challenge of controlling the release and rate of degradation, which is usually investigated by conducting \textit{in vitro} and \textit{in vivo} tests requiring conducting multiple experiments for different scenarios and situations. In this study, we have developed a mathematical model to predict the biodegradation behavior of commercially pure Mg-based biomaterials, which makes it possible to study the corrosion of implants and scaffolds in a simulated environment. Despite the assumed simplifications, the model can serve as an important tool to find the biodegradable metals properties and predict the biodegradation behavior of Mg-based implants that improves current design workflows.

\section*{Acknowledgment}

This research is financially supported by the Prosperos project, funded by the Interreg VA Flanders – The Netherlands program, CCI grant no. 2014TC16RFCB046 and by the Fund for Scientific Research Flanders (FWO), grant G085018N. LG acknowledges support from the European Research Council under the European Union’s Horizon 2020 research and innovation programme, ERC CoG 772418. The computational resources and services used in this work were provided by the VSC (Flemish Supercomputer Center), funded by the Research Foundation - Flanders (FWO) and the Flemish Government – department EWI.

\bibliographystyle{elsarticle-num}
\bibliography{mybibs}
\end{document}